\documentclass[a4paper,usenatbib, 14pt]{jpconf}
\usepackage{graphicx,amsmath,iopams}
\usepackage{color} 

\newcommand{\Mmax}{m_\mathrm{max}}
\newcommand{\Mmin}{m_\mathrm{min}}
\newcommand{\vcrit}{v_\mathrm{crit}}
\newcommand{\NeRat}{$^{22}$Ne/$^{20}$Ne}
\newcommand{\Nfour}{$^{14}$N}

\newcommand{\NeTT}{$^{22}$Ne}

\begin{document}

\title{Wolf-Rayet stars in young massive star clusters as potential sources of  Galactic cosmic rays}%

\author{M E Kalyashova$^{1}$,  A M Bykov$^{1,2}$, S M Osipov$^{1}$, D C Ellison$^{3}$,  D V Badmaev$^{1}$}
\address{$^1$ Ioffe Institute, 26 Politekhnicheskaya st., St. Petersburg 194021, Russia}
\address {$^2$ Peter The Great St. Petersburg Polytechnic University, 29 Politekhnicheskaya st., St. Petersburg 195251, Russia}
\address {$^3$ Department of Physics, North Carolina State University,  Raleigh, NC 27695-8202, USA}

\ead{filter-happiness@yandex.ru}

\begin{abstract}
For most elements, the isotopic ratios seen in cosmic rays (CRs) are similar to those in the solar wind.
The most important exception to this is \NeRat\ where the CR value is $\sim 5$ times that of the solar wind.
According to most recent models of nucleosynthesis, a large amount of $^{22}\mathrm{Ne}$ is generated in Wolf-Rayet (WR) stars. In the winds of carbon sequence WR stars, i.e., WC stars, the isotopic ratio $^{22}\mathrm{Ne}$/$^{20}\mathrm{Ne}$  can be much larger than in the solar wind. 
Here, we consider CRs produced by $^{22}\mathrm{Ne}$-enriched WR winds in compact clusters of young massive stars like Westerlund~1. 
We assume that  efficient CR acceleration in clusters occurs over the WR lifetime in an ensemble of shock waves from multiple massive star winds.
We estimate the fraction of all Galactic CRs such sources may produce for a given set of parameters.
\end{abstract}

\section{Introduction}

The origin of   Galactic cosmic rays (GCRs) is a long-standing problem of the theoretical and observational astrophysics. 
From composition (e.g., \cite{MDE97}) and energy budget requirements (e.g., \cite{GS64}), the main sources of CRs with energies below the knee at $\sim 10^{15}$\,eV are likely supernova remnants (SNRs), where particles are accelerated at collisionless shocks due to the 
first-order Fermi mechanism (also called diffusive shock acceleration).
%
However, other galactic sources, such as pulsars and stellar winds, are certain to contribute at some level. Strong stellar winds from massive stars are a likely source (see e.g. \cite{CP80, CM81,BT81,vf82,Bykov01,Binns08,Lingen2018,Seo2018}) and this acceleration will be enhanced in compact star clusters of massive stars where acceleration can take place by the ensemble of shocks from multiple winds \cite{Bykov2014,Aharonian2019}.

The chemical composition of GCRs has been investigated with several experiments (IMP-7\cite{Garcia1979}, ISEE-3\cite{Wiedenbeck1981}, Voyager\cite{Lukasiak1994}, ACE-CRIS\cite{Binns2005} and others), while the solar wind composition was deeply studied by determining meteoritic CI chondrite abundance (see, e.g.,\cite{Lodders2003}). It has been shown that abundances of most elements and isotopes are similar to solar-system abundances once the enhanced contribution of high-mass elements and refractory elements from interstellar dust are considered 
(see \cite{EDM97}). 
The most important isotope difference is the \NeRat\  ratio.
In the solar wind 
$^{22}\mathrm{Ne}$/$^{20}\mathrm{Ne}=0.07$, while in GCRs $^{22}\mathrm{Ne}$/$^{20}\mathrm{Ne}=0.387 \pm 0.027$, which corresponds to $^{22}\mathrm{Ne}$ enhancement by a factor of $5.3 \pm 0.3$  \cite{Binns2005}.

After the discovery of the $^{22}\mathrm{Ne}$ overabundance, a number of possible mechanisms of this phenomena were proposed 
(e.g., \cite{Woosley1981, Reeves1978, Olive1982}). In 1982 Casse and Paul \cite{Casse1982} introduced their explanation of the neon excess, which is currently the most widely accepted one. 
{They suggested that the overabundance of \NeTT\ came from ejecta of carbon stage Wolf-Rayet (WC) stars.
During He-burning in WR stars almost all \Nfour\ is transformed into
$^{22}\mathrm{Ne}$ through the chain of reactions $^{14}\mathrm{N}(\alpha, \gamma)$ $ ^{18}\mathrm{F}(e^+ \nu) $ $^{18}\mathrm{O} (\alpha, \gamma)$ $^{22}\mathrm{Ne}$.
%
%
The $^{22}\mathrm{Ne}$-enriched material is then expelled in the stellar wind of the WC star. Casse and Paul estimated that the $^{22}\mathrm{Ne}$ excess at the surface of WC stars is $\sim 120$, so, the contribution of such sources to the GCRs should be $\sim 2 \%$. This mechanism was later studied quantitatively using the numerical models of massive stars \cite{Prantzos1987, MaederMeynet1993}. 

While WR stars seem a likely source of \NeTT, the mechanism of particle acceleration remains an issue. CR acceleration may occur in galactic superbubbles produced by  OB star associations with multiple supernovae \cite{Bykov2001, Lingen2018}. Higdon and Lingenfelter (2003) \cite{Higdon2003} suggested that superbubbles, where the combined action of SN shock waves and stellar winds takes place, can be the possible sources of  a substantial fraction of GCRs. Using the stellar models by Schaller et al.~\cite{Schaller1992}, they found that the observed value of $^{22}\mathrm{Ne}$/$^{20}\mathrm{Ne}$ can be achieved with a mixture of $\sim 20 \%$ WR star material and $\sim 80 \%$ material with standard composition. Another idea, proposed by Prantzos~\cite{Prantzos2012}, is that GCRs are accelerated by the forward shocks of SN explosions as they run through the pre-supernova winds of the massive stars and through the interstellar medium.  Prantzos suggested that acceleration takes place in the Sedov-Taylor phase of the SNR.

In this work we suggest that young massive star clusters can be a significant source of GCRs and enhanced \NeRat. It is considered that energetic particles can be effectively accelerated by very energetic outflows with multiple shocks from massive star winds. We examine a typical cluster and show the dependency of $^{22}\mathrm{Ne}$/$^{20}\mathrm{Ne}$  on cluster age and the cluster initial mass function (IMF). We take one of the most well investigated massive clusters, Westerlund~1, as an example. 
The fraction of massive stars born in compact clusters compared to those born in  loose OB-association is still debated \cite{Ward2018}.
For the point of our interest, we consider both 
compact clusters with scales below a parsec, and larger OB associations at early stages of their evolution, to show similar effects contributing  to enhanced \NeRat\ in CRs.
%

\section{Particle acceleration at colliding shock flows in massive star clusters}

For more than three decades the interacting winds of massive stars have been considered as potential CR accelerators \cite{CP80,Axford81,BT81,Casse1982,vf82,cm83,EU93}. 
Particle acceleration in SN shock waves and massive star winds in OB associations and superbubbles was studied in detail in \cite{Bykov2001,Ferrand2010, Bykov2014}.
It was shown recently in  \cite{Bykov2014,BEOG15} that a supernova occurring in a  compact cluster of young massive stars like  
Westerlund~1 can accelerate  CRs to well above PeV energies with an acceleration time below $\sim 10^3$\,yr.
%
While these events can produce observable PeV neutrinos fluxes, during most of their lifetime they will accelerate lower energy CRs
in shocks and MHD turbulence produced by the colliding winds of the very young stars in the compact cluster.
%
Kinetic equations describing  the collective action of a system of many supernova shocks were derived in \cite{BT1990,Klepach2000}.     

Spectroscopic data show that WR stars, along with O- and B-type stars, have powerful stellar winds with velocities of $1000-3000$ km/s. 
Due to their small size and high star density \cite{Zwart2010},  young massive star clusters are likely to have  strong colliding magnetohydrodynamic shock flows, where efficient particle acceleration can take place.
Recently, Seo et al.~\cite{Seo2018} estimated the total power of O- and WR stars' winds in the Galaxy as $ \sim 1.1 \times 10^{41}$ erg/s.
If $(1-10) \%$ of the wind luminosity is converted to GCR energy, these  stellar winds can provide a significant contribution to the GCR production below the knee.
 
It is an open issue how many stars in the Galaxy are members of star clusters. While it is often assumed that most stars are formed in dense star clusters, there is little agreement on the precise fraction (see, e.g., \cite{Ward2018}). Future surveys (e.g. \textit{Gaia}) are expected to clarify the situation.
The aim of our current work is to investigate the role compact clusters have in the production and acceleration of CR \NeTT.

\section{Method}

To get the neon isotopic yields, as well as lifetimes of massive stars and mass loss rates, we use one of the most recent stellar evolution models of the Geneva group (Ekstr{\"o}m et al.~\cite{Ekstrom2012}, 
Georgy et al.~\cite{Georgy2012}). They have calculated grids of non-rotating and rotating models with metallicity $Z=0.014$ and initial star masses from 0.8 to 120 $M_{\odot}$. 
The velocity of rotation in rotating models is $v=0.4  \vcrit$, where $\vcrit =\sqrt{2 GM/3R}$ ($G$ is the gravitational constant, $M$ is the star mass, $R$ is the star polar
radius)~\cite{Ekstrom2012}. 
To determine the mass of neon isotopes ejected by O-, B- and WR stars, we use interpolations of the Geneva group tabulations of the neon surface mass fractions, $\Sigma_{22} (m, t)$ and $\Sigma_{20}(m, t)$, the mass loss rate $\dot{M}(m, t)$, and the lifetime of the star $t_l$. Thus, for a single star with the initial mass $m$, the total mass of neon isotope $i$ ($i=20, 22$), ejected in the wind by the time $t$, equals:
\begin{equation}
\mathcal{M}_{i}(m, t)=\int_{0}^{t} \Sigma_{i}(m, t') \dot{M}(m, t') dt' 
\ .
\end{equation}
If $t>t_l$, then $\mathcal{M}_{i}(m, t)=\mathcal{M}_{i}(m, t_l)$. For the entire cluster, we sum the isotopic yields $\mathcal{M}_{i}$ folded with the initial mass function (IMF), $\chi(m)=dn/dm$,
where $\chi(m)dm$ is the number of stars per unit volume with the initial mass between $m$ and  $m+dm$. The IMF in different environments is studied in experiments and usually is represented in the power-law form:
\begin{equation}
\chi(m)=A \cdot m^{-\gamma}    
\ .
\end{equation}
Taking this into account, the neon isotopic ratio at a given moment $t$ (i.e. the ratio of ejected masses of $^{22}\mathrm{Ne}$ and $^{20}\mathrm{Ne}$ from the whole cluster by the time $t$) equals:
\begin{equation}
    \frac{^{22}\mathrm{Ne}}{^{20}\mathrm{Ne}}= 
    \frac{\int^{\Mmax}_{\Mmin} \mathcal{M}_{22}(m,t) m^{-\gamma} dm}
    {\int^{\Mmax}_{\Mmin} \mathcal{M}_{20}(m,t) m^{-\gamma} dm}
\ .
\end{equation}
For our calculation we use the range of initial star masses from $\Mmin =15 M_{\odot}$ to $\Mmax =120 M_{\odot}$.

The power-law index $\gamma$ is usually determined as $\gamma \approx 2.3-2.7$~\cite{Seo2018, Kroupa2002, Salpeter1955}. 
However, recent studies revealed that some clusters (e.g. Arches\cite{Hosek2019} and Westerlund~1\cite{Lim2013}) may have a much flatter IMF with $\gamma \approx 1.8$. In order to examine the impact of the cluster IMF on the neon isotopic ratio,  we use three different indices in our calculations: $\gamma=1.8$, $2.3$, and $2.6$.

\section{Results and discussion}

In Figure 1 we present the $^{22}\mathrm{Ne}$/$^{20}\mathrm{Ne}$ isotopic ratio for a single cluster as a function of the cluster age, assuming different IMF power-law indices. An inspection of Figure 1 shows that:
\begin{itemize}
\item [--] Both rotating and non-rotating models give a maximum of \NeRat\  at 4-5 Myr. At these times, the amount of ejected $^{22}\mathrm{Ne}$ is significantly larger than $^{20}\mathrm{Ne}$.
\item [--] The isotopic ratio decreases with time and becomes constant after $\sim 12$ Myr. In all cases, that constant value is larger than observed in GCRs.
\item [--] The isotopic ratio increases with the IMF flattening, because the fraction of massive stars which can become a WR star also increases.
\item [--] Generally, rotating models give more $^{22}\mathrm{Ne}$ than non-rotating ones. This can be explained with the fact that rotation allows stars with smaller initial masses ($\sim 30 M_{\odot}$) to become WR stars~\cite{Georgy2012}.
\end{itemize}

We have obtained results showing that star clusters can begin producing a substantial amount of \NeTT\ in CRs after $\sim 3$\,Myr. 
Over their lifetime, we estimate they may contribute at least 
$30-50$\% of the galactic CR enrichment in \NeRat,  depending on the number of  clusters.
The remaining fraction of  \NeTT\ enriched CRs may be accelerated during the  
early evolutionary stages of OB associations.
%
%
%
One  of the most massive and well-known clusters is Westerlund~1. Considering its age ($\sim 5$\,Myr), its  IMF power-law index 
($\gamma \sim 1.8$), and the fact that  observations
(e.g. \cite{Wd1_Fenech}) show plenty of WC  stars, it is a prime candidate for producing CR \NeTT.
%
%
Of course, keeping in mind the residence time of GeV CRs in the Galaxy 
($>10$ Myr), it is impossible to point to particular clusters as sites of \NeTT\ enrichment.
Our results suggest that compact massive star clusters may be a significant source of  the observed neon isotopic ratio. 
 
\begin{figure}[h]
\begin{center}
\includegraphics[scale=0.8]{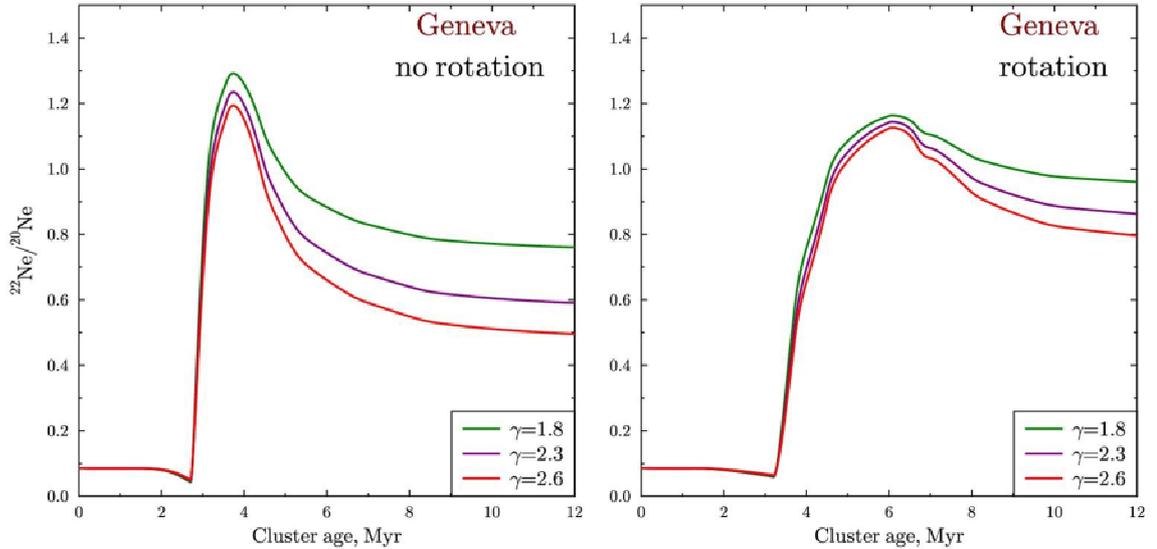}
\end{center}
\caption{Left panel: neon isotopic ratio as a function of star cluster age for three values (as indicated) of the IMF power-law index for 
stellar models \textit{without rotation}. Right panel: the same for 
stellar 
models \textit{with rotation}.}
\label{nrr}
\end{figure}

\section{Summary}
We examine young massive star clusters as an alternative source of  Galactic CRs and, in particular, the observed \NeRat\ ratio. 
We assume that efficient particle acceleration takes place in the multiple shock waves produced by interacting winds of O-, B-, and WR stars.
%
Using state-of-the-art  stellar evolution models, we calculate the amount of neon isotopes produced by such sources and conclude that massive star clusters can produce  a significant fraction of GCRs and \NeTT.



\ack
A.M.B. acknowledges support from RAS Presidium program No. 12.
The results of the work were obtained using computational resources of Peter the Great Saint-Petersburg Polytechnic University Supercomputing Center (http://www.spbstu.ru)




\section*{References}

\bibliographystyle{iopart-num}
\bibliography{bibliogr}
\end{document}